\documentclass[preprint,12pt]{elsarticle}

\usepackage{amssymb}

\usepackage{amsmath,amsthm,amssymb,mathrsfs}
\usepackage{bm}

\journal{Journal of Magnetism and Magnetic Materials}

\begin{document}

\begin{frontmatter}




\title{Synchronized, periodic, and chaotic dynamics in spin torque oscillator with two free layers}


\author{Tomohiro Taniguchi 
}


\address{
 National Institute of Advanced Industrial Science and Technology (AIST), Spintronics Research Center, Tsukuba, Ibaraki 305-8568, Japan, 
}



\begin{abstract}
A phase diagram of the magnetization dynamics is studied by numerically solving the Landau-Lifshitz-Gilbert (LLG) equation 
in a spin torque oscillator consisting of asymmetric two free layers that are magnetized in in-plane direction.  
We calculated the dynamics for a wide range of current density for both low and high field cases, 
and found many dynamical phases such as synchronization, auto-oscillation with different frequencies, and chaotic dynamics. 
The observation of the synchronization indicates the presence of a dynamical phase which has not been found experimentally 
by using the conventional electrical detection method. 
The auto-oscillations with different frequencies lead to an oscillation of magnetoresistance with a high frequency, 
which can be measured experimentally. 
The chaotic and/or periodic behavior of magnetoresistance in a high current region, on the other hand, leads to a discontinuous change of the peak frequency in Fourier spectrum. 
\end{abstract}


\end{frontmatter}





\section{Introduction}
\label{sec:Introduction}

Current driven magnetization dynamics, such as switching and oscillation, in nanostructured ferromagnetic multilayers 
have been attractive research topics in the fields of spintronics, nonlinear science, and applied physics 
\cite{slonczewski89,slonczewski96,berger96,katine00,kiselev03,kubota05,hillebrands06,krivorotov07,krivorotov08,bertotti09,rippard10,kubota13}. 
The switching dynamics is used as an operation principle of magnetic random access memory (MRAM) \cite{dieny16}, 
whereas spin torque oscillator (STO) utilizing an auto-oscillation (limit cycle) of the magnetization can be used as an element for
microwave generator, magnetic sensor, phased array radar, or brain-inspired computing \cite{locatelli14,grollier16,torrejon17,kudo17,tsunegi18SR,romera18}. 
A ferromagnetic multilayer used in spintronics devices usually includes three ferromagnets called free, reference, and pinned layers \cite{dieny16}. 
Spin currents polarized by the reference layer are injected into the free layer and excite the magnetization dynamics in the free layer via spin transfer torque effect \cite{slonczewski96,berger96,slonczewski05}. 
Note that the spin transfer occurs not only in the free layer but also in the reference layer. 
Therefore, a finite spin torque is simultaneously acting on the magnetization in the reference layer. 
Efforts have been made to clarify the role of spin torque on the magnetization dynamics in the reference layer \cite{gusakova09,gusakova11,kudo12,matsumoto14,tsunegi18}. 
The magnetization in the reference layer, however, has usually been assumed to be fixed 
because an antiferromagnetic coupling from the pinned layer strongly pins the magnetization. 


Recently, however, there is a motivation to study the magnetizations dynamics in two ferromagnets coupled via spin transfer effect. 
The development in the magnetic recording technology faces a serious issue 
because the magnetic field produced in conventional recording method solely using a direct field is not sufficient enough in next generation high-density recording system. 
Microwave assisted magnetization reversal (MAMR) is a new scheme of magnetic recording, 
where the microwave emitted from an STO contributes to the reduction of the direct magnetic field necessary for the recording 
\cite{bertotti01,thirion03,zhu08,okamoto12,suto12,kudo14,taniguchi14PRB,suto15,taniguchi16PRB,suto17,suto18}. 
The latest design of the STO for MAMR consists of two in-plane magnetized ferromagnetic layers called field-generation layer and spin-injection layer \cite{comment1}. 
The field-generation layer acts as a microwave generator for MAMR, whereas the spin-injection layer is the source of spin current injected into the field-generation layer. 
Importantly, this type of STO does not include a pinned layer in order to make the recording head small for high density recording \cite{comment1,comment2}, 
and therefore, both two ferromagnets can be regarded as free layers. 
Experimental efforts investigating the oscillation properties of this type of STO have been reported very recently \cite{comment2,comment4}. 
Magnetization dynamics possibly occurring in this type of STO is, however, not fully understood yet. 


Magnetization dynamics in an STO is studied by using a giant magnetoresistive (GMR) or tunneling magnetoresistance (TMR) effect, 
where an oscillating voltage (or resistance) output from the multilayer depends on the relative angle between the free and reference layers. 
For an STO with a pinned layer \cite{kiselev03,krivorotov07,krivorotov08,rippard10,kubota13}, 
an electrical signal produced from these magnetoresistance effects directly reflects the magnetization dynamics in the free layer 
because the magnetization in the reference layer can be assumed to be fixed. 
On the other hand, for an STO with two free layers, the electrical signal does not necessarily reflect the magnetization dynamics. 
For example, let us for the moment assume that the magnetizations in two free layers show a synchronization, which is typically observed in a coupled system \cite{pikovsky03}. 
In this case, no oscillating electrical signal can be obtained from the STO because the relative angle between the magnetization is constant as a function of time. 
The dynamics is, however, applicable to a microwave generator 
because the magnetizations in the free layers are oscillating, and as a result, microwave field is emitted from the STO. 
As can be understood from this example, there might be a hidden magnetization dynamics in this type of STO, which has not been clarified experimentally \cite{comment2,comment4}. 
To overcome this issue, a theoretical analysis will be useful. 



In this paper, we study the magnetization dynamics in a ferromagnetic multilayer consisting of two in-plane magnetized free layers 
by solving the Landau-Lifshitz-Gilbert (LLG) equation numerically. 
Several dynamical phases, such as synchronization, auto-oscillation with different frequencies, and chaotic dynamics, are found over a wide range of applied electric current. 
The synchronization of the auto-oscillations having identical frequency appears in a low current region. 
This result clarifies the existence of a dynamical phase which cannot be detected by the conventional electrical measurement. 
The auto-oscillations with different frequencies, on the other hand, can be measured because the magnetoresistance shows an oscillation with high frequency. 
In the highly nonlinear periodic and/or chaotic dynamical phase which appeared in a high current region, 
the Fourier spectrum shows multipeaks with small amplitudes due to the complex dynamics. 
The oscillation (peak) frequency cannot be well-defined in this region. 
The electrical detection of the chaotic motion however is possible because such a dynamics leads to a discontinuous change of the peak frequency in the Fourier spectrum. 




\begin{figure}
\centerline{\includegraphics[width=0.5\columnwidth]{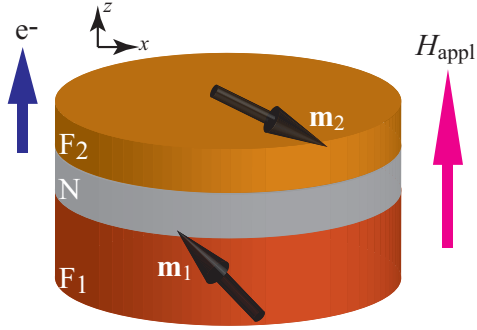}}
\caption{
        Schematic illustration of system used in this study. 
        A nonmagnet (N) is sandwiched by two ferromagnetic layers F${}_{k}$ ($k=1,2$)
        The unit vector pointing in the magnetization direction of F${}_{k}$ layer is denoted as $\mathbf{m}_{k}$. 
        The $z$ axis is perpendicular-to-plane direction. 
        The external magnetic field $H_{\rm appl}$ is applied to the positive $z$ direction. 
        The positive current corresponds to the electrons flowing from F${}_{1}$ to F${}_{2}$ layer. 
         \vspace{-3ex}}
\label{fig:fig1}
\end{figure}




\section{System description}
\label{sec:System description}

In this section, we describe the system employed in this study and show the explicit form of the LLG equation. 


\subsection{LLG equation}
\label{sec:LLG equation}

The system we consider is schematically shown in Fig. \ref{fig:fig1}, 
where two ferromagnets F${}_{k}$ ($k=1,2$) sandwiches a nonmagnet N. 
The unit vector pointing in the magnetization direction of F${}_{k}$ layer is denoted as $\mathbf{m}_{k}$. 
In the following, similar to $\mathbf{m}_{k}$, we add the suffix $k$ to quantities related to F${}_{k}$ layer. 
The electric current density $j$ flows along the $z$ direction, 
where a positive current corresponds to the electrons flowing from F${}_{1}$ to F${}_{2}$ layer. 
The LLG equation of F${}_{k}$ layer is given by \cite{slonczewski96,slonczewski05} 
\begin{equation}
\begin{split}
  \frac{d \mathbf{m}_{k}}{dt}
  =&
  -\gamma
  \mathbf{m}_{k}
  \times
  \mathbf{H}_{k}
  +
  \alpha_{k}
  \mathbf{m}_{k}
  \times
  \frac{d \mathbf{m}_{k}}{dt}
\\
  &-
  \frac{\gamma \hbar p_{k}j}{2e(1+p_{k}^{2}\mathbf{m}_{1}\cdot\mathbf{m}_{2})M_{k}d_{k}}
  \mathbf{m}_{k}
  \times
  \left(
    \mathbf{m}_{2}
    \times
    \mathbf{m}_{1}
  \right), 
  \label{eq:LLG}
\end{split}
\end{equation}
where $\gamma$ and $\alpha_{k}$ are the gyromagnetic ratio and the Gilbert damping constant, respectively. 
The magnetic field $\mathbf{H}_{k}$ consists of an applied field $H_{\rm appl}$, 
the demagnetization field, and the dipole field as 
\begin{equation}
  \mathbf{H}_{k}
  =
  \begin{pmatrix}
    -4\pi M_{k} N_{kx} m_{kx} - H_{{\rm d}k} m_{k^{\prime}x} \\
    -4\pi M_{k} N_{ky} m_{ky} - H_{{\rm d}k} m_{k^{\prime}y} \\
    H_{\rm appl} - 4\pi M_{k} N_{kz} m_{kz} + 2 H_{{\rm d}k} m_{k^{\prime}z}
  \end{pmatrix}. 
\end{equation}
The saturation magnetization of F${}_{k}$ layer is denoted as $M_{k}$. 
The demagnetization coefficient $N_{ki}$ ($i=x,y,z$) 
and the dipole field $H_{{\rm d}k}$ acting from F${}_{k^{\prime}}$ to F${}_{k}$ layer ($k^{\prime} \neq k$)
in a cylindrically shaped multilayer are, respectively, given by \cite{tandon03,taniguchi18JMMM}, 
\begin{equation}
  N_{kz}
  =
  \frac{1}{\tau_{k}}
  \left\{
    \frac{4}{3\pi}
    -
    \frac{4}{3\pi}
    \sqrt{
      1
      +
      \tau_{k}^{2}
    }
    \left[
      \tau_{k}^{2}
      \mathsf{K}
      \left(
        \frac{1}{\sqrt{1+\tau_{k}^{2}}}
      \right)
      +
      \left(
        1
        -
        \tau_{k}^{2}
      \right)
      \mathsf{E}
      \left(
        \frac{1}{\sqrt{1+\tau_{k}^{2}}}
      \right)
    \right]
    +
    \tau_{k}
  \right\},
\end{equation}
\begin{equation}
  H_{{\rm d}k}
  =
  \pi
  M_{k^{\prime}}
  \left[
    \frac{\frac{d_{k}}{2}+d_{\rm N}+d_{k^{\prime}}}{\sqrt{r^{2}+\left( \frac{d_{k}}{2}+d_{\rm N}+d_{k^{\prime}} \right)^{2}}}
    -
    \frac{\frac{d_{k}}{2}+d_{\rm N}}{\sqrt{r^{2}+\left( \frac{d_{k}}{2}+d_{\rm N} \right)^{2}}}
  \right],
\end{equation}
where $\tau_{k}=d_{k}/(2r)$. 
The thicknesses of F${}_{k}$ and N layers are denoted as $d_{k}$ and $d_{\rm N}$, respectively, 
whereas the radius of the multilayer is $r$. 
Because of the assumption of the cylindrical shape, $N_{kx}=N_{ky}=(1-N_{kz})/2$. 
The first and second kind of complete elliptic integrals with a modulus $\Bbbk$ are denoted as $\mathsf{K}(\Bbbk)$ and $\mathsf{E}(\Bbbk)$, respectively. 
The spin polarization characterizing the strength of the spin torque acting on the magnetization in F${}_{k}$ layer is denoted as $p_{k}$. 
In our definition, the spin torque acting on F${}_{1}$ (F${}_{2}$) layer excited by a positive current prefers 
an antiparallel (parallel) alignment of the magnetizations. 
This point will be used to understand the coupled motions of the magnetizations studied in the following sections. 


\subsection{Values of parameters}
\label{sec:Values of parameters}

The values of the parameters used in this paper are estimated from typical experiments focusing on the oscillation behavior of an STO for MAMR 
consisting of CoFe/Ag/NiFe multilayer \cite{comment2,comment3}, where CoFe and NiFe correspond to F${}_{1}$ and F${}_{2}$ layers, respectively. 
That is, 
$M_{1}=1720$ emu/c.c., $M_{2}=800$ emu/c.c., 
$\alpha_{1}=0.006$, $\alpha_{2}=0.010$, 
$d_{1}=5$ nm, $d_{2}=3$ nm, $r=50$ nm, $d_{\rm N}=5$ nm, 
$p_{1}=p_{2}=0.3$, and $\gamma=1.764 \times 10^{7}$ rad/(Oe s). 
Using these values, we find that 
$N_{1z}=0.876$, $N_{2z}=0.916$, $H_{{\rm d}1}=143.7$ Oe, and $H_{{\rm d}2}=514.6$ Oe. 
The value of the radius $r$ in this work is assumed to be larger than the experimental value ($15$ nm) \cite{comment2} so that the macrospin model becomes applicable. 
In the present model, F${}_{1}$ and F${}_{2}$ layers correspond to the field-generation and spin-injection layers, respectively. 
Since the multilayer consists of metals, it is experimentally possible to apply a large current on the order of $10^{8}$ A/cm${}^{2}$ \cite{comment2}, 
contrary to an STO using a magnetic tunnel junction \cite{hiramatsu16}, 
where the maximum value of the current density is on the order of $10^{6}-10^{7}$ A/cm${}^{2}$. 


There are some previous works on a coupled motion of the magnetizations in two free layers in a multilayer \cite{urazhdin08,kudo06,kani16,kani17}. 
In these, two ferromagnets have almost identical properties 
because the works principally focused on a synchronized oscillation of the magnetizations or simultaneous switching of two magnetizations, 
although chaotic motion was also observed \cite{kudo06}. 
On the other hand, the parameters of F${}_{1}$ and F${}_{2}$ in the present work are largely different from each other due to an asymmetric structure of the STO for MAMR \cite{comment1,comment2,kanai17,kanai18}. 
In MAMR, the field-generation layer should have large saturation magnetization and volume to emit sufficient strength of microwave field to recording media 
whereas the spin-injection layer should be thin to make small sized recording head. 
In particular, the factor $1/(M_{2}d_{2})$ is almost four times larger than $1/(M_{1}d_{1})$, 
and therefore, the spin torque acting on F${}_{2}$ layer is much larger than that acting on F${}_{1}$ layer. 
As a result, not only the synchronized motion of the magnetizations but also other kinds of the magnetization dynamics, such as a chaotic behavior, will be observed in the present system, as shown below. 


The value of the applied field $H_{\rm appl}$ used in the following calculation is chosen to be $6.0$ or $20.0$ kOe. 
We call $H_{\rm appl}=6.0$ ($20.0$) kOe the low (high) field. 
These values are chosen to satisfy $H_{\rm appl}<4\pi M_{1},4\pi M_{2}$ and $H_{\rm appl}>4\pi M_{1},4\pi M_{2}$ 
for the low and high field cases, respectively. 
We first solve the LLG equation without the spin torque terms, 
and chose the relaxed state as an initial condition. 
For the low field case, the initial states of the magnetizations are deviated from the $z$ axis, i.e., the magnetizations are not parallel to the applied field. 
For the high field case, on the other hand, the relaxed sates of the magnetizations are parallel to the applied field along the $z$ axis. 
The details of the initial conditions are summarized in \ref{sec:AppendixA}. 


\section{Phase diagram}
\label{sec:Phase diagram}

An auto-oscillation of an STO has been studied by measuring the relation between the current and oscillation frequency \cite{kiselev03,krivorotov07,rippard10,kubota13}. 
In particular, for MAMR application, the microwave frequency from the STO is of interest, 
which corresponds to the oscillation frequency of the in-plane ($xy$ plane) component of the magnetization. 
Therefore, we evaluate the peak frequency of the Fourier spectrum of $m_{kx}$ in the following. 
In addition, we remind the readers that the auto-oscillation of the magnetization has been measured through the GMR or TMR effect \cite{kiselev03,krivorotov07,krivorotov08,rippard10,kubota13}, 
where the resistance of a multilayer is related to the magnetizations $\mathbf{m}_{1}$ and $\mathbf{m}_{2}$ via \cite{slonczewski89,bauer03} 
\begin{equation}
  R
  =
  \frac{R_{\rm AP}+R_{\rm P}}{2}
  -
  \frac{R_{\rm AP}-R_{\rm P}}{2}
  \mathbf{m}_{1}
  \cdot
  \mathbf{m}_{2},
\end{equation}
where $R_{\rm P}$ and $R_{\rm AP}$ are the resistance of the system 
in the parallel and antiparallel alignment of the magnetizations, respectively. 
Therefore, in the following, we call the vector product between $\mathbf{m}_{1}$ and $\mathbf{m}_{2}$ as MR, for convention,
\begin{equation}
  {\rm MR}
  \equiv
  \mathbf{m}_{1}
  \cdot
  \mathbf{m}_{2},
\end{equation}
and investigate not only the dynamics of $\mathbf{m}_{1}$ and $\mathbf{m}_{2}$ but also that of the MR. 



\begin{figure}
\centerline{\includegraphics[width=1.0\columnwidth]{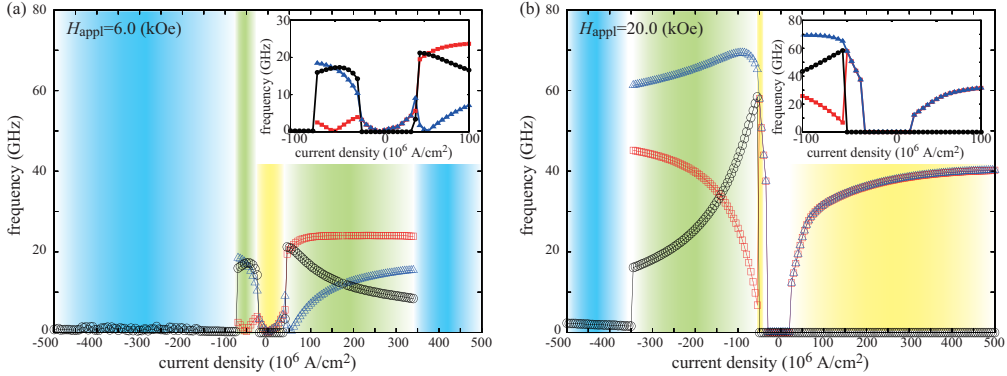}}
\caption{
        Dependences of the oscillation (peak) frequencies of F${}_{1}$ (red square), F${}_{2}$ (blue triangle), and the MR (black circle) on the applied current density 
        for (a) the low ($H_{\rm appl}=6.0$ kOe) and (b) the high ($H_{\rm appl}=20.0$ kOe) field cases. 
        The yellow, green, and blue-shaded regions correspond to the synchronized oscillation region, 
        the auto-oscillations with different frequencies region, and chaotic region, respectively. 
        A peak frequency in the chaotic region is not well-defined. 
        Near zero current for $H_{\rm appl}=20.0$ kOe, an instability threshold appears in addition, in which the magnetizations cannot move from the equilibrium states. 
        The insets show the current-frequency relation for the current range of $-100.0 \times 10^{6} \le j \le 100.0 \times 10^{6}$ A/cm${}^{2}$. 
         \vspace{-3ex}}
\label{fig:fig2}
\end{figure}




The oscillation (peak) frequencies of $m_{1x}$, $m_{2x}$, and the MR for the low and high field cases are summarized in Figs. \ref{fig:fig2}(a) and \ref{fig:fig2}(b), respectively. 
The details of the magnetization dynamics shown in these figures will be described in Secs. \ref{sec:Low field case} and \ref{sec:High field case}, respectively. 
Note here that, in some regions, an oscillation (peak) frequency is not well-defined, and thus, not shown in the figure. 
For example, when a chaotic dynamics is excited, an instantaneous frequency varies as a function of time. 
For these cases, the oscillation frequencies cannot be defined, and therefore, are excluded from Figs. \ref{fig:fig2}(a) and \ref{fig:fig2}(b). 
Also, when a synchronization occurs, MR is constant, and thus, the peak frequency of MR is zero. 
These dynamics will also be described in the following sections. 



\section{Low field case}
\label{sec:Low field case}

In this section, we show the current dependence of the magnetization dynamics in the low field case. 
The corresponding phase diagram is depicted Fig. \ref{fig:fig2}(a). 
Since the dynamic behavior changes drastically depending on the current magnitude, we briefly summarize the results beforehand. 
In the low current region, a synchronized oscillation of the magnetizations appears. 
In the middle current region, two magnetizations show auto-oscillations with different frequencies. 
In the high current region, the magnetization dynamics becomes chaotic. 




\begin{figure}
\centerline{\includegraphics[width=1.0\columnwidth]{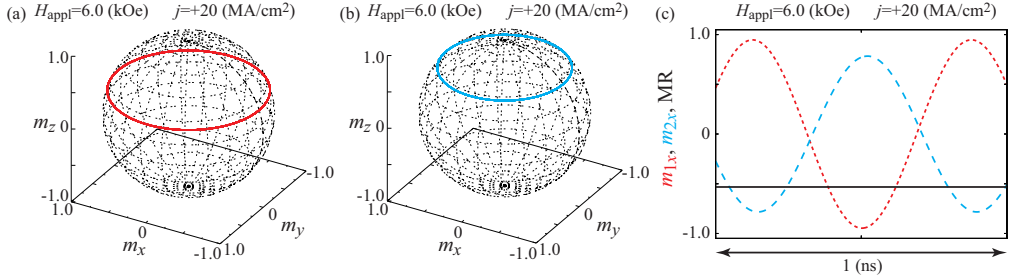}}
\caption{
        The oscillation trajectories of (a) $\mathbf{m}_{1}$ and (b) $\mathbf{m}_{2}$ 
        and (c) the time evolutions of $m_{1x}$ (red dotted), $m_{2x}$ (blue dashed), and the MR (black solid) 
        at $j=+20.0 \times 10^{6}$ A/cm${}^{2}$ and $H_{\rm appl}=6.0$ kOe. 
         \vspace{-3ex}}
\label{fig:fig3}
\end{figure}



\subsection{Synchronized oscillations in low current region}
\label{sec:Synchronized oscillations in low current region}

In this section, we show the magnetization dynamics excited in the low current region with a low applied field, corresponding to the yellow-shaded region in Fig. \ref{fig:fig2}(a). 
The trajectories of the magnetization dynamics in F${}_{1}$ and F${}_{2}$ layers typically found in this current region 
are shown in Figs. \ref{fig:fig3}(a) and Fig. \ref{fig:fig3}(b), respectively. 
The time evolutions of $m_{1x}$, $m_{2x}$, and the MR are shown in Fig. \ref{fig:fig3}(c) by the red dotted, blue dashed, and black solid lines, respectively. 
The current density is $j=+20.0 \times 10^{6}$ A/cm${}^{2}$. 
The auto-oscillations around the $z$ axis are excited in both layers. 
In addition, the oscillation frequencies of two magnetizations are identical, i.e., a synchronization is excited. 
As a result, the MR is constant as a function of time, as shown in Fig. \ref{fig:fig3}(c). 
Therefore, the oscillation frequency of the MR is zero in the low current region, 
and the STO does not generate an electrical signal. 

The phase synchronization between mutually coupled STOs has been observed experimentally \cite{tsunegi18SR,kaka05,mancoff05,sani13,locatelli15,urazhdin16,houshang16,awad17}. 
In the experiments, an in-phase synchronization is easy to measure because the in-phase synchronization leads to the enhancement of an output power \cite{tsunegi18SR}. 
On the other hand, an antiphase synchronization is predicted theoretically \cite{slavin09,turtle13,kendziorczyk14,taniguchi17,turtle17,taniguchi18APEX,taniguchi18PRB1,taniguchi18PRB2},  
depending on the coupling mechanism and/or structures, 
although the antiphase synchronization is difficult to detect experimentally because output power in this case becomes nearly zero. 
In a symmetric system including two auto-oscillators, either in-phase or antiphase synchronization is stable, depending on the coupling mechanism \cite{pikovsky03}. 
Contrary to the results of these previous works, the phase difference between the synchronized magnetizations in the present system is neither in-phase nor antiphase, as can be seen in Fig. \ref{fig:fig3}(c). 
This is because two ferromagnets have different free-running frequencies \cite{pikovsky03}. 
In addition, the coupling via spin torque is asymmetric; for example, 
for a positive current, the spin torque acting on F${}_{1}$ layer acts as a repulsive force to F${}_{2}$ layer, 
whereas the torque acting on F${}_{2}$ layer is an attractive force to F${}_{1}$ layer. 
As a result, the phase difference between two magnetizations neither converge to the in-phase nor to the antiphase state. 

We should emphasize that, although an electrical detection of the oscillations of the magnetizations through the magnetoresistance effect is difficult due to the synchronization, 
the magnetizations in this low current region show auto-oscillations, and may be able to be applied to the microwave source of MAMR. 




\begin{figure}
\centerline{\includegraphics[width=1.0\columnwidth]{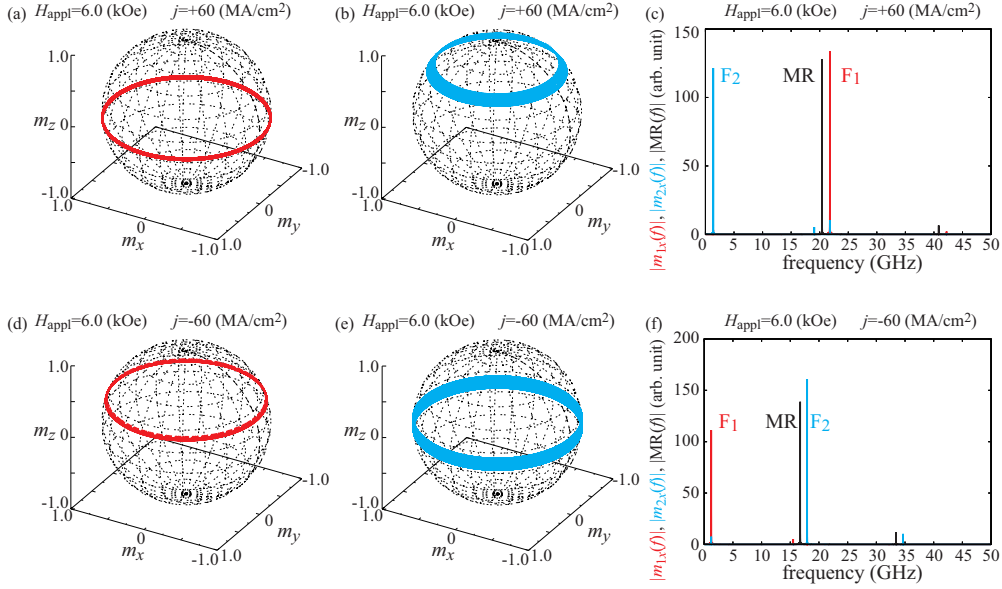}}
\caption{
        Oscillation trajectories of (a) $\mathbf{m}_{1}$ and (b) $\mathbf{m}_{2}$ 
        and (c) the Fourier transformations of $m_{1x}$ (red dotted), $m_{2x}$ (blue dashed), and the MR (black solid) 
        at $j=+60.0 \times 10^{6}$ A/cm${}^{2}$ and $H_{\rm appl}=6.0$ kOe. 
        Those for a negative current $j=-60.0 \times 10^{6}$ A/cm${}^{2}$ are shown in (d)-(f). 
         \vspace{-3ex}}
\label{fig:fig4}
\end{figure}



\subsection{Oscillations in middle current region}
\label{sec:Oscillations in middle current region}

This section shows the magnetization dynamics excited in the middle current region with a low applied field, corresponding to the green-shaded region in Fig. \ref{fig:fig2}(a). 
Figures \ref{fig:fig4}(a)-(c) and \ref{fig:fig4}(d)-(f) show the oscillation trajectories of the magnetizations 
and their Fourier transformations for a positive ($j=+60 \times 10^{6}$ A/cm${}^{2}$) and negative ($j=-60 \times 10^{6}$ A/cm${}^{2}$) current densities, respectively. 
The red, blue, and black lines in Figs. \ref{fig:fig4}(c) and \ref{fig:fig4}(f) are $|m_{1x}(f)|$, $|m_{2x}(f)|$, and the Fourier transformation of the MR, respectively. 
In this region, two magnetizations oscillate with different frequencies. 
As a result, the oscillation frequency of the MR is the difference or sum of the oscillation frequencies of $\mathbf{m}_{1}$ and $\mathbf{m}_{2}$, depending on the oscillating directions. 

Recall that the initial states of the magnetizations are close to the positive $z$ direction because the external field is applied to this direction. 
Since the spin torque acting on the magnetization in F${}_{1}$ excited by a positive current prefers the antiparallel alignment of the magnetizations, 
$\mathbf{m}_{1}$ moves to the negative $z$ region by the spin torque, whereas $\mathbf{m}_{2}$ remains in the positive $z$ region, as can be seen in Figs. \ref{fig:fig4}(a) and \ref{fig:fig4}(b).
Note that the oscillation frequency around the $z$ axis is determined by the effective field, which includes a term $H_{\rm appl}-4\pi M_{k}m_{kz}$. 
Since $m_{1z}<0$ and $m_{2z}>0$, the effective field of F${}_{1}$ layer is larger than that of F${}_{2}$ layer. 
Therefore, the oscillation frequency of F${}_{1}$ layer is higher than that of F${}_{2}$ layer, as shown in Fig. \ref{fig:fig4}(c). 
When the current direction is changed to the negative direction, the spin torques acting on two ferromagnets reverse their directions. 
Thus, for a negative current case, the oscillation frequency of F${}_{1}$ becomes lower than that of F${}_{2}$ layer, as shown in Fig. \ref{fig:fig4}(f). 






\begin{figure}
\centerline{\includegraphics[width=1.0\columnwidth]{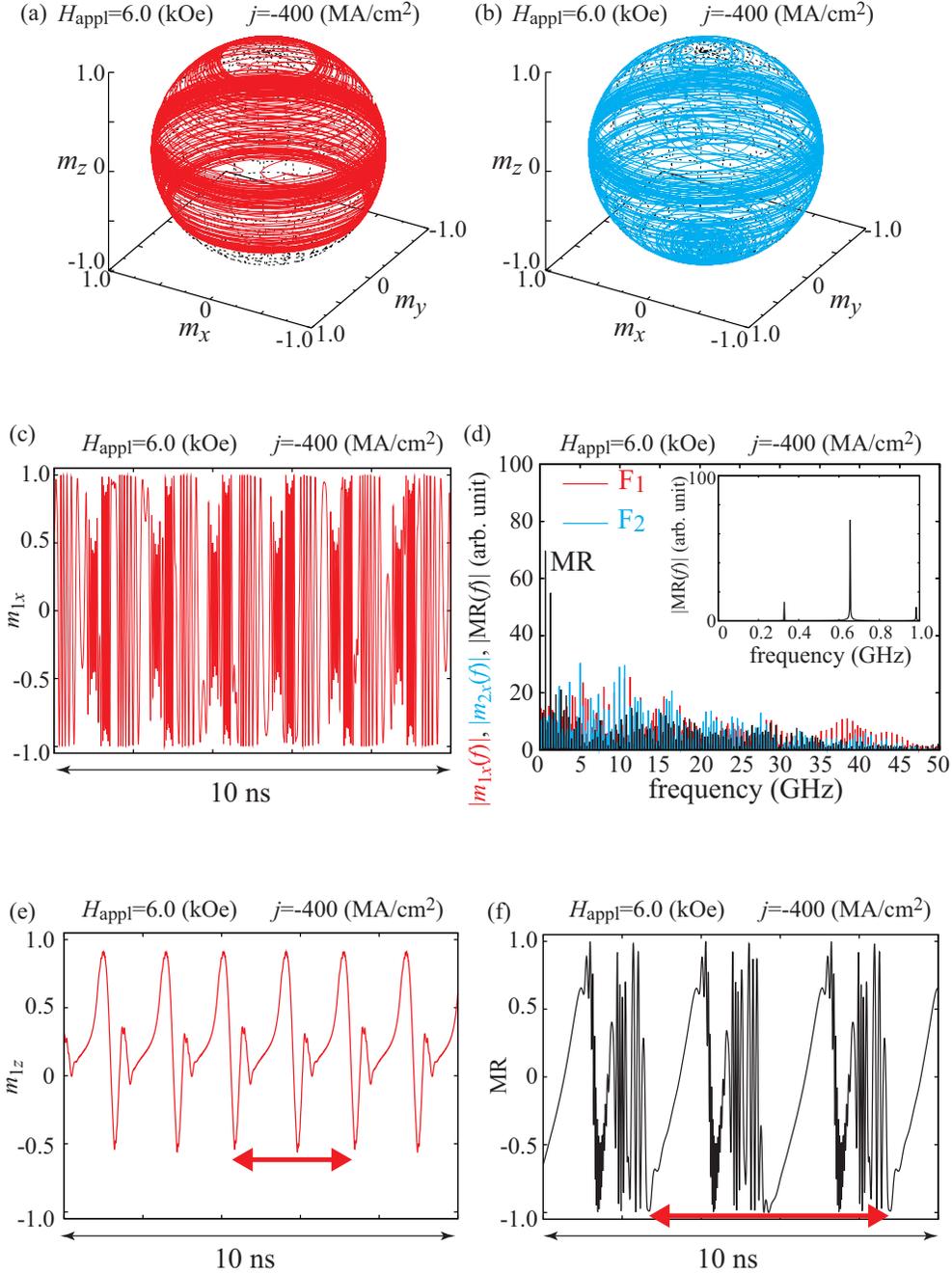}}
\caption{
        Oscillation trajectories of (a) $\mathbf{m}_{1}$ and (b) $\mathbf{m}_{2}$ at $H_{\rm appl}=6.0$ kOe and $j=-400 \times 10^{6}$ A/cm${}^{2}$. 
        (c) Time evolution of $m_{1x}$. 
        (d) Fourier transformations of $m_{1x}$ (red), $m_{2x}$ (blue), and MR (black). 
            The inset shows the Fourier transformation of the MR in the low frequency region. 
        (e) Time evolution of $m_{1z}$. The red allow corresponds to the period of $m_{1z}$. 
        (f) Time evolution of the MR. The red allow corresponds to the period of the MR. 
         \vspace{-3ex}}
\label{fig:fig5}
\end{figure}



\subsection{Chaotic dynamics of magnetizations and periodic MR in high current region (negative current)}
\label{sec:Chaotic dynamics of magnetizations and periodic MR in high current region (negative current)}

In this section, we show the magnetization dynamics excited in the high and negative current region with a low applied field, corresponding to the blue-shaded left region in Fig. \ref{fig:fig2}(a). 
Figures \ref{fig:fig5}(a) and \ref{fig:fig5}(b) show the trajectories of $\mathbf{m}_{1}$ and $\mathbf{m}_{2}$ at $j=-400 \times 10^{6}$ A/cm${}^{2}$, respectively. 
Since the strength of the spin torque is sufficiently large due to a large current magnitude, 
the magnetizations widely change their directions, as shown in these figures.
Figure \ref{fig:fig5}(c) shows the time evolution of the in-plane component, $m_{1x}$, of the magnetization in F${}_{1}$ layer. 
Although $m_{1x}$ repeats a similar oscillation pattern, we did not observe any periodicity in a strict sense as shown in Fig. \ref{fig:fig5}(c) or even in a widen time range. 
The Fourier transformation of $m_{1x}$ shown in Fig. \ref{fig:fig5}(d) has many peaks with small amplitudes; see, for example, Fig. \ref{fig:fig4}(c) for comparison. 
The peak (oscillation) frequency of such a spectrum cannot be well-defined as well. 
The Fourier transformation of $m_{2x}$ also shows many small and comparable peaks, 
and therefore, a peak frequency is not well-defined. 
The absence of the periodicity in the nonlinear dynamics of $m_{1x}$, 
as well as the sensitive dependence of the dynamics on its initial condition shown in \ref{sec:AppendixB}, 
indicates that its dynamics is a chaos. 
Recall that the chaos appears due to the presence of the coupling between two ferromagnets via spin-transfer effect and dipole field, 
whereas it is precluded for a single ferromagnet with the macrospin model because of the Poincar\'e-Bendixson theorem \cite{bertotti09,bertotti07,strogatz01}. 
We note that chaos in an STO is also found in Ref. \cite{comment5}, where a delayed feedback is used to induce the chaos. 

On the other hand, the perpendicular component of the magnetization, $m_{1z}$, shows a periodic motion, as shown in Fig. \ref{fig:fig5}(e), where the period is shown by the red arrow in the figure. 
In addition, the MR($=\mathbf{m}_{1}\cdot\mathbf{m}_{2}$) also shows a periodic motion, as shown in Fig. \ref{fig:fig5}(f). 
Although the Fourier transformation of the MR also shows multipeaks, there is a large peak in the low frequency region, as can be seen in Fig. \ref{fig:fig5}(d). 
The fact that the MR varies from $-1$ to $+1$ means that the magnetization alignment changes between the parallel and antiparallel alignment periodically. 
This is because the spin torque acting on F${}_{1}$ layer prefers the parallel alignment whereas that on F${}_{2}$ layer prefers the antiparallel alignment. 

Summarizing these results, the magnetization dynamics in each ferromagnet projected to the film-plane shows a chaotic behavior. 
The Fourier spectrum shows multipeaks with small amplitudes over a wide range of the frequency. 
On the other hand, the MR shows a periodic behavior, and the Fourier spectrum shows a sharp peak at low frequency region. 
Therefore, only the peak frequency of the MR is plotted in Fig. \ref{fig:fig2}(a). 
Recall that, in the middle current region, the peak frequency of the MR was relatively high because of the stable oscillations of the magnetizations. 
Thus, the low frequency of the MR found in the high current region indicates that 
a discontinuous drop of the MR peak frequency at the boundary between the middle and high current regions 
is expected in the experimental measurement, as can be seen in the negative current region in Fig. \ref{fig:fig2}(a). 
We also note that this current region cannot be applied to MAMR application because the microwave frequency from the STO is not fixed. 




\begin{figure}
\centerline{\includegraphics[width=1.0\columnwidth]{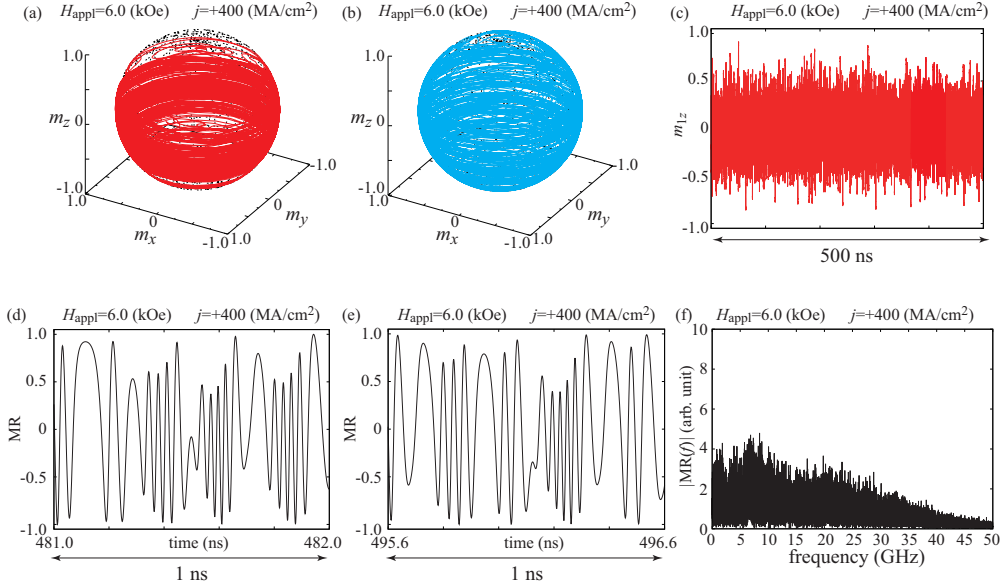}}
\caption{
        Oscillation trajectories of (a) $\mathbf{m}_{1}$ and (b) $\mathbf{m}_{2}$ at $H_{\rm appl}=6.0$ kOe and $j=+400.0 \times 10^{6}$ A/cm${}^{2}$. 
        (c) Time evolution of the MR with the time range of 10 ns. 
        (d), (e) Examples of the time evolution of the MR with the time range of 1 ns. 
        (f) Fourier transformation of the MR. 
         \vspace{-3ex}}
\label{fig:fig6}
\end{figure}



\subsection{Chaotic behavior of magnetizations and MR in high current region (positive current)}
\label{sec:Chaotic behavior of magnetizations and MR in high current region (positive current)}

Next, we show the magnetization dynamics excited in the high and positive current region with a low applied field, corresponding to the blue-shaded right region in Fig. \ref{fig:fig2}(a). 
Figures \ref{fig:fig6}(a) and \ref{fig:fig6}(b) show the dynamic trajectories of $\mathbf{m}_{1}$ and $\mathbf{m}_{2}$. 
Similar to the negative current region discussed in the previous section, 
the magnetizations widely change their directions due to the large spin torque. 
Contrary to the negative current case, however, not only the in-plane components of the magnetizations 
but also the perpendicular components show chaotic behavior. 
For example, the time evolution of $m_{1z}$ shown in Fig. \ref{fig:fig6}(c) does not show a periodicity over a wide time range (500 ns). 
Figures \ref{fig:fig6}(d) and \ref{fig:fig6}(e) show examples of the time evolution of the MR, where the time range in both figures is 1 ns. 
These figures indicate that the time evolution of the MR repeats a similar oscillation pattern, but it is not periodic. 
As a result, not only the magnetizations but also the MR does not have sharp peaks in their Fourier transformations: 
see, for example, Fig. \ref{fig:fig6}(f), where the Fourier transformation of the MR is shown. 
Thus, no peak frequency is well-defined in this current region, indicating the absence of periodicity. 
In addition, this fact means that no visible (or strong) peak of emission power will be observed experimentally. 
Therefore, symbols are not shown in the high positive current region in Fig. \ref{fig:fig2}(a). 


\subsection{Suggestions to verify synchronization}

As mentioned in Sec. \ref{sec:Synchronized oscillations in low current region}, an electrical detection of the synchronization through GMR or TMR effect is difficult 
because the magnetoresistance is constant as a function of time in this case. 
In this section, let us discuss alternative approaches to verify the synchronization experimentally. 
The problem is that we cannot distinguish 
whether the magnetizations are in auto-oscillation states with synchronization or at the state of pointing to fixed directions without oscillations 
from the GMR or TMR effect in the present geometry. 
Therefore, to verify the synchronization, it is sufficient to detect an auto-oscillation of, at least, one magnetization. 

The first suggestion is to use an electrical detection. 
For example, when another ferromagnet having a fixed in-plane magnetization is placed on F${}_{2}$ layer in Fig. \ref{fig:fig1}, 
the GMR or TMR effect between this additional ferromagnet and F${}_{2}$ layer provides an oscillating electric voltage. 
Such a multilayer structure having three ferromagnets was proposed in Ref. \cite{kent04} for a different purpose. 
An alternative approach for the electrical detection is to place a Hall bar below, for example, F${}_{1}$ layer in Fig. \ref{fig:fig1}. 
Applying electric current to the Hall bar, anisotropic magnetoresistance (AMR) \cite{mcguire75} 
and/or spin Hall magnetoresistance (SMR) \cite{nakayama13,althammer13,kim16} effects show an oscillating behavior reflecting the auto-oscillation in F${}_{1}$ layer. 
It should, however, be noted that adding another ferromagnet and/or Hall bar provide 
additional spin torques related to GMR \cite{slonczewski96,berger96}, TMR \cite{slonczewski89}, AMR \cite{taniguchi15PRApplied}, and/or SMR \cite{chen13,chiba_thesis} effects, which might disturb the synchronization. 

The second suggestion is to measure MAMR in another ferromagnet located near the STO. 
Recently, MAMR assisted by microwave generated from an in-plane magnetized STO was experimentally demonstrated in magnetic nanodot placed on the STO \cite{suto16}. 
Using a similar method, the existence of microwave field is verified, which is an evidence of an auto-oscillation in the STO. 

The reader might be interested not only in the verification of the synchronization but also in its relaxation phenomenon 
because the relaxation time is related to, for example, the recording speed of MAMR. 
In \ref{sec:AppendixC}, we show examples of the relaxation phenomenon in the present STO. 


\section{High field case}
\label{sec:High field case}

In this section, we show the current dependence of the magnetization dynamics in the high field region. 
The corresponding phase diagram is Fig. \ref{fig:fig2}(b). 
As shown below, the magnetization dynamics is similar to that observed in the low field case. 
There are, however, differences, such as an existence of an instability threshold. 


\subsection{Threshold current}
\label{sec:Threshold current}

A difference between the low and high field cases is the existence of an instability threshold for the high field case, 
i.e., non-zero current is necessary to excite the magnetization dynamics. 
Recall that the synchronized auto-oscillation for the low field case 
described in Sec. \ref{sec:Synchronized oscillations in low current region} was excited for the current density of $|j|>0$.
On the other hand, for the high field case studied below, threshold current densities to excite magnetization dynamics exist for both the positive and negative current densities. 
For the present system, when the current density $j$ is in the range of $-30.0 \times 10^{6} \le j \le +20.0 \times 10^{6}$ A/cm${}^{2}$, 
the magnetizations do not move from the initial equilibrium state even in the presence of the spin torque. 

The existence of instability threshold of an in-plane magnetized free layer has been investigated previously \cite{grollier03,firastrau07,houssameddine07,ebels08,silva10,taniguchi16PRB2,taniguchi18JJAP}, 
where the magnetization in the reference layer is fixed. 
Although it is difficult to extend these works to the two free layers system analytically, 
the existence of the instability threshold is qualitatively explained by using these single free layer models, as explained below. 

The instability threshold appears due to the damping torque, which prevents the magnetization to move from the energetically stable state to a high energy state. 
In the absence of the external field, the energetically stable state of the free layer is the in-plane magnetized state. 
Note that the magnetic energy does not change even when the magnetization changes its direction in the film plane, i.e., the energetically minimum state forms a constant energy surface in the film plane. 
In such a case, a spin torque with any magnitude can move the magnetization within the film plane because the damping torque does not prevent such motion. 
Therefore, the instability threshold current is zero. 
This conclusion still holds even in the presence of the external field, 
if the field magnitude is smaller than the demagnetization field. 
On the other hand, when the field magnitude is larger than the demagnetization field, the energetically stable state becomes parallel to the $z$ axis, i.e., the energetically minimum state is a point. 
In this case, a finite energy injection as a work done by spin torque is necessary to move the magnetization to any direction 
by overcoming the dissipation due to the damping torque. 
Therefore, the instability threshold current becomes finite. 

Regarding the result shown in Sec. \ref{sec:Synchronized oscillations in low current region}, 
we come to the conclusion that no electric power can be obtained near the zero current for both low and high field cases. 
We should emphasize, however, that there is a difference between the two. 
For the low field case, the electrical signal does not appear due to the synchronization, although each magnetization itself shows auto-oscillation. 
On the other hand, for the high field case, no movement at all for the magnetizations is observed from the equilibrium state. 
The former is applicable to practical applications such as a microwave generator, whereas the latter is not.


\subsection{Negative current region}
\label{sec:Negative current region}



\begin{figure}
\centerline{\includegraphics[width=1.0\columnwidth]{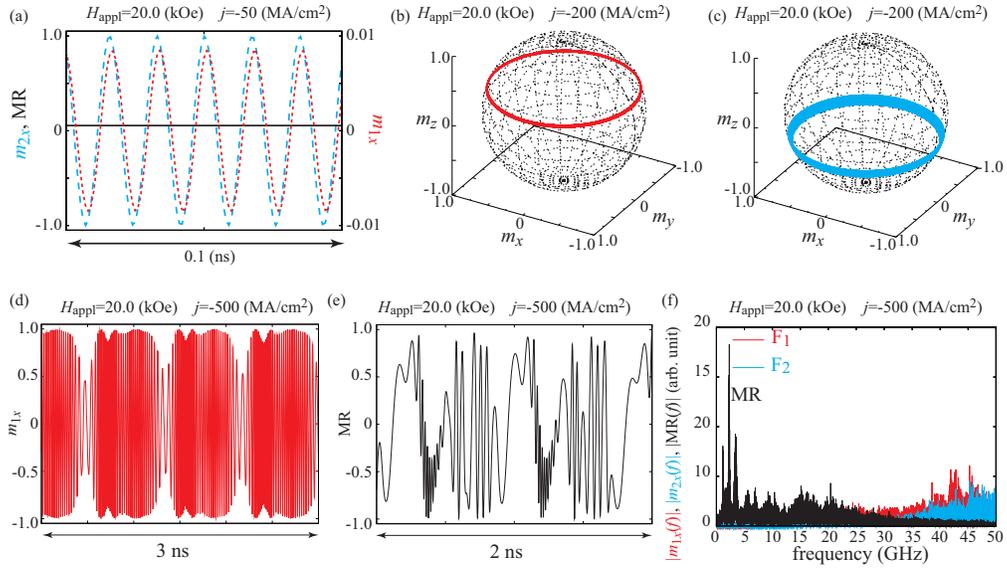}}
\caption{
        (a) Time evolutions of $m_{1x}$ (red dotted), $m_{2x}$ (blue dashed), and the MR (black solid) at $H_{\rm appl}=20.0$ kOe and $j=-50.0 \times 10^{6}$ A/cm${}^{2}$.
        Oscillation trajectories of (b) $\mathbf{m}_{1}$ and (c) $\mathbf{m}_{2}$ at $H_{\rm appl}=20.0$ kOe and $j=-200.0 \times 10^{6}$ A/cm${}^{2}$. 
        Time evolutions of (d) $m_{1x}$ and (e) the MR at $H_{\rm appl}=20.0$ kOe and $j=-500.0 \times 10^{6}$ A/cm${}^{2}$. 
        (f) Fourier transformations of $m_{1x}$ (red), $m_{2x}$ (blue), and the MR (black) at $H_{\rm appl}=20.0$ kOe and $j=-500.0 \times 10^{6}$ A/cm${}^{2}$.
         \vspace{-3ex}}
\label{fig:fig7}
\end{figure}



Below the threshold ($j=-30.0\times 10^{6}$ A/cm${}^{2}$), the magnetization dynamics similar to that found in Sec. \ref{sec:Low field case} is excited by the negative current. 
For $-50.0 \times 10^{6} \le j \le -35.0 \times 10^{6}$ A/cm${}^{2}$, a synchronization of the auto-oscillations in two ferromagnets is excited. 
Figure \ref{fig:fig7}(a) shows an example of the synchronization, where the magnetizations $\mathbf{m}_{1}$ and $\mathbf{m}_{2}$ oscillate with the same frequency, 
whereas the MR is constant as a function of time. 
When the magnitude of the current density increases ($-345.0 \times 10^{6} \le j \le -55.0 \times 10^{6}$ A/cm${}^{2}$), 
the synchronization disappears, and the magnetizations oscillate with different frequencies. 
Figures \ref{fig:fig7}(b) and \ref{fig:fig7}(c) show examples of the dynamic trajectories of $\mathbf{m}_{1}$ and $\mathbf{m}_{2}$ for such case. 
Since the spin torque acting on F${}_{2}$ layer prefers the antiparallel alignment of the magnetizations, $\mathbf{m}_{2}$ moves to the negative $z$ region. 
In a large current density limit ($j \le-350.0 \times 10^{6}$ A/cm${}^{2}$), a chaotic dynamics of the magnetizations appears. 
Figures \ref{fig:fig7}(d) and \ref{fig:fig7}(e) show examples of the time evolutions of $m_{1x}$ and the MR, respectively, where $j=-500.0 \times 10^{6}$ A/cm${}^{2}$. 
These quantities show non-periodic oscillations. 
In addition, these dynamics are sensitive to the initial states; see \ref{sec:AppendixB}. 
The Fourier spectrum of the MR shows a large peak at a certain frequency, 
whereas the spectra of F${}_{1}$ and F${}_{2}$ layers have multipeaks with small amplitudes, as can be seen in Fig. \ref{fig:fig7}(f). 


\subsection{Positive current region}
\label{sec:Positive current region}



\begin{figure}
\centerline{\includegraphics[width=1.0\columnwidth]{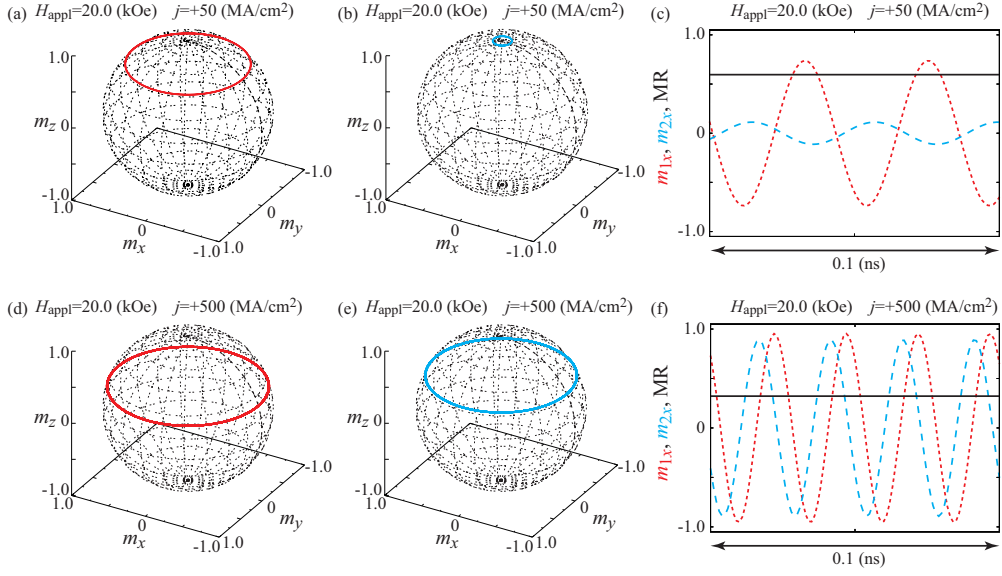}}
\caption{
        Oscillation trajectories of (a) $\mathbf{m}_{1}$ and (b) $\mathbf{m}_{2}$ at $H_{\rm appl}=20.0$ kOe and $j=+50.0 \times 10^{6}$ A/cm${}^{2}$. 
        (c) Time evolutions of $m_{1x}$ (red dotted), $m_{2x}$ (blue dashed), and the MR (black solid) with the time range of 0.1 ns.  
        Similar plots for $j=+500.0 \times 10^{6}$ A/cm${}^{2}$ are shown in (d)-(f). 
         \vspace{-3ex}}
\label{fig:fig8}
\end{figure}



In the positive current region above the threshold ($j=+20.0 \times 10^{6}$ A/cm${}^{2}$), 
a synchronized auto-oscillation of the magnetizations is found over a wide range of current. 
Figures \ref{fig:fig8}(a)-(c) and \ref{fig:fig8}(d)-(f) show 
the oscillation trajectories of the magnetizations and the time evolutions of $m_{1x}$, $m_{2x}$, and the MR for 
$j=+50.0 \times 10^{6}$ A/cm${}^{2}$ and $j=+500.0 \times 10^{6}$ A/cm${}^{2}$, respectively. 
Recall that the synchronization also appeared in the low field case. 
However, the range of the current density corresponding to the synchronization for the present (high field) case is significantly larger than that for the low field case. 
This is because the high field stabilizes the oscillation around the $z$ axis.




\section{Conclusion}
\label{sec:Conclusion}

In conclusion, a phase diagram of the magnetization dynamics in a spin torque oscillator consisting of asymmetric two in-plane magnetized free layers was derived 
by solving the Landau-Lifshitz-Gilbert (LLG) equation.  
Several dynamical phases, classified as synchronization, auto-oscillation with different frequencies, and chaotic behavior, were found in low and high field cases. 
In addition, an instability threshold appeared in the high field case. 
The study revealed the presence of the synchronization in the low current region, 
although the synchronized auto-oscillations will be difficult to detect by conventional electrical detection method. 
The auto-oscillations with different frequencies, on the other hand, lead to the oscillation of the magnetoresistance in the STO, and thus, can be measured in experiments. 
In a sufficiently large current limit, the magnetization dynamics showed a chaotic behavior. 
Nevertheless, the magnetoresistance, which is proportional to the vector product of two magnetizations, showed a periodic oscillation, depending on the current and field conditions. 
Theoretical demonstrations of such rich varieties of coupled dynamics introduced in this study will provide a deep insight on nonlinear dynamics in nanostructures 
and may contribute to designing practical devices such as microwave generators.


\section*{Acknowledgment}
The author express the deepest gratitude to Hitoshi Kubota, Shingo Tamaru, Takehiko Yorozu, Sumito Tsunegi, Weinan Zhou, and Yuya Sakuraba for valuable discussions. 
The author is also thankful to Satoshi Iba, Aurelie Spiesser, Hiroki Maehara, and Ai Emura for their support and encouragement. 






\appendix


\section{Details of numerical simulation}
\label{sec:AppendixA}

In the numerical simulation, we first solve the LLG equation without spin torque to determine the energetically stable states of the magnetizations. 
In this calculation, the magnetizations are initially set as $\mathbf{m}_{1}=+\mathbf{e}_{x}$ and $\mathbf{m}_{2}=-\mathbf{e}_{x}$. 
Then, the magnetizations finally saturate to energetically stable states, which are found to be 
$\mathbf{m}_{1}=(0.84979,0.39435,0.34978)$ and $\mathbf{m}_{2}=(-0.67025,-0.31103,0.67381)$ for the low field case. 
We use these values as the initial states in the presence of the spin torque. 
We should note that these states have a rotational symmetry, i.e., the magnetic energy does not change even when these states are rotated around the $z$ axis. 
On the other hand, for the high field case, the stable states become $\mathbf{m}_{1},\mathbf{m}_{2}=+\mathbf{e}_{z}$. 
Note that all torques, including the spin torque, in the LLG equation become zero in this magnetization alignment. 
Therefore, we slightly shift the initial states of the magnetizations as 
$\mathbf{m}_{k}=(\sqrt{1-m_{kz}^{2}},0,1-(k \times 10^{-5}))$ to produce non-zero torque acting on the magnetization.


\section{Sensitive dependence on initial condition in chaotic phase}
\label{sec:AppendixB}

The chaos is a dynamical phase in which the long-term prediction of the dynamics is impossible \cite{strogatz01}. 
In this context, a periodic motion is not a chaos. 
For example, although the dynamics of the magnetization shown in Fig. \ref{fig:fig5}(c) repeats a similar pattern, it is not periodic in a strict sense. 
The identification of the chaos, however, also requires to study the sensitivity to the initial state \cite{strogatz01}. 
When a dynamical trajectory is in an attractor in a phase space, 
even if it looks complicated, the dynamics with close but slightly different initial conditions finally converges to an identical trajectory. 
Such a dynamics is easy to predict once a trajectory at a certain initial condition is clarified. 



\begin{figure}
\centerline{\includegraphics[width=1.0\columnwidth]{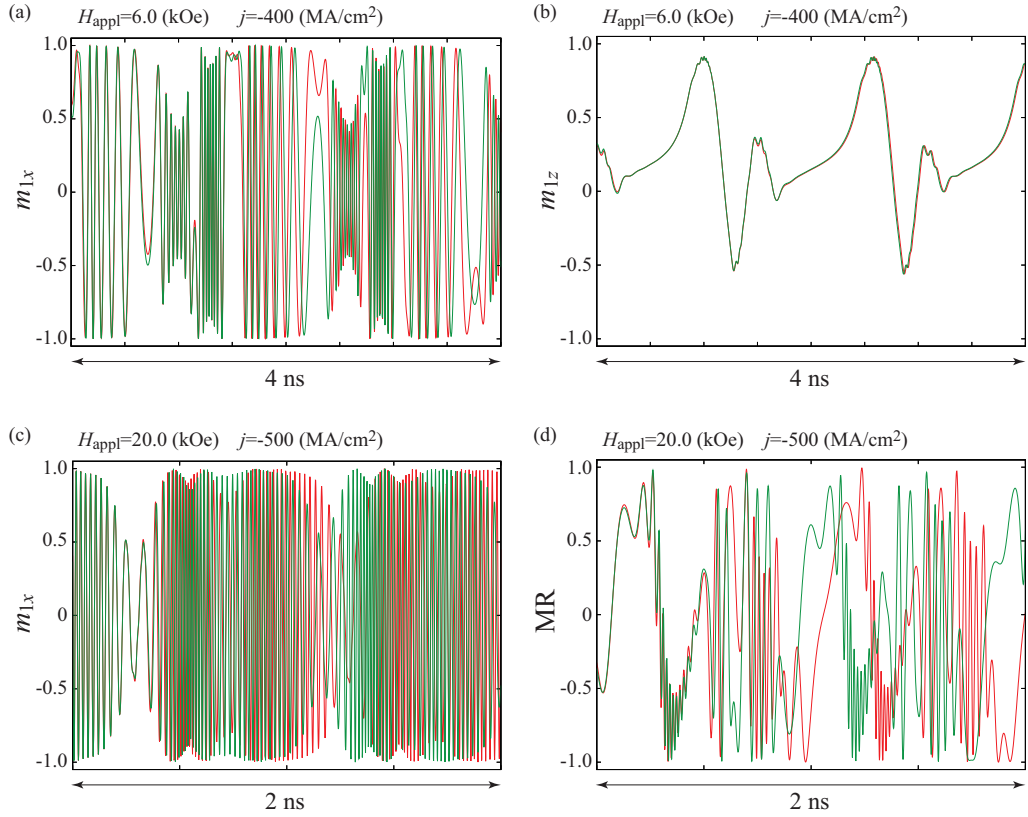}}
\caption{
        Time evolutions of (a) $m_{1x}$ and (b) $m_{1z}$ with slightly different values at $t=490$ ns. 
        The magnetic field and current density are $H_{\rm appl}=6.0$ kOe and $j=-400.0 \times 10^{6}$ A/cm${}^{2}$, respectively. 
        Red lines in (a) and (b) are identical to those in Figs. \ref{fig:fig5}(c) and \ref{fig:fig5}(e). 
        Time evolutions of (c) $m_{1x}$ and (d) the MR in the high field ($H_{\rm appl}=20.0$ kOe) case with $j=-500.0 \times 10^{6}$ A/cm${}^{2}$ are also shown. 
         \vspace{-3ex}}
\label{fig:fig9}
\end{figure}



The sensitivity of the magnetization dynamics in Fig. \ref{fig:fig5}(c) is studied as follows. 
In Fig. \ref{fig:fig9}(a), the red line is $m_{1x}(t)$ at $H_{\rm appl}=6.0$ kOe and $j=-400 \times 10^{6}$ A/cm${}^{2}$, where $490 \le t \le 494$ ns. 
Recall that the red line is obtained with the initial conditions given in \ref{sec:AppendixA}. 
To study whether this dynamics is a chaos or in an attractor, 
we slightly change the value of $\mathbf{m}_{1}$ at $t=490$ ns. 
Explicitly, we change the value as $m_{1x}^{\prime}(t=490\ {\rm ns})=a m_{1x}(t=490\ {\rm ns})$, 
$m_{1y}^{\prime}(t=490\ {\rm ns})=b m_{1y}(t=490\ {\rm ns})$, and $m_{1z}^{\prime}(t=490\ {\rm ns})=b m_{1z}(t=490\ {\rm ns})$, 
where a coefficient $a$ is chosen to be 0.9, 
whereas $b$ is $\sqrt{[1-a^{2} m_{1x}^{2}(t=490\ {\rm ns})]/[1-m_{1x}^{2}(t=490\ {\rm ns})]}$ to keep the norm of $\mathbf{m}_{1}^{\prime}=(m_{1x}^{\prime},m_{1y}^{\prime},m_{1z}^{\prime})$ one. 
The green line shown in Fig. \ref{fig:fig9}(a) is $m_{1x}^{\prime}$, which has a slightly different value to $m_{1x}$ at $t=490$ ns. 
As shown, near $t=490$ ns, $m_{1x}$ and $m_{1x}^{\prime}$ show similar dynamics. 
Soon after, however, a difference between $m_{1x}$ and $m_{1x}^{\prime}$ appears, in particular when the instantaneous frequency is small. 
In other words, $m_{1x}$ and $m_{1x}^{\prime}$ do not converge to an identical trajectory. 
The result means that the dynamics of $m_{1x}$ is sensitive to the initial condition. 
Therefore, we conclude in the main text that the dynamics of $m_{1x}$ is chaos. 

On the other hand, the dynamics of $m_{1z}$ shown in Fig. \ref{fig:fig5}(e) is not a chaos because the dynamics shows periodicity. 
Another evidence that the dynamics of $m_{1z}$ not being chaos can be seen in Fig. \ref{fig:fig9}(b), 
where $m_{1z}$ and $m_{1z}^{\prime}$ are shown by the red and green lines, respectively. 
As shown, starting from the different states, $m_{1z}$ and $m_{1z}^{\prime}$ finally converge to the identical trajectory, 
which does not satisfy the definition of the chaos. 


The sensitivity to the initial state for the high field ($H_{\rm appl}=20.0$ kOe) case is also studied. 
Figures. \ref{fig:fig9}(c) and \ref{fig:fig9}(d) show 
the time evolutions of $m_{1x}$ and the MR with slightly different conditions at $t=490$ ns, 
where the current density is $-500.0 \times 10^{6}$ A/cm${}^{2}$, as in the case shown in Figs. \ref{fig:fig7}(d) and \ref{fig:fig7}(e). 
The dynamics of both the magnetization and the MR can be observed to depend on the initial conditions. 
Therefore, in Sec. \ref{sec:Negative current region}, we concluded that the magnetizations and the MR show chaotic behavior. 




\begin{figure}
\centerline{\includegraphics[width=0.8\columnwidth]{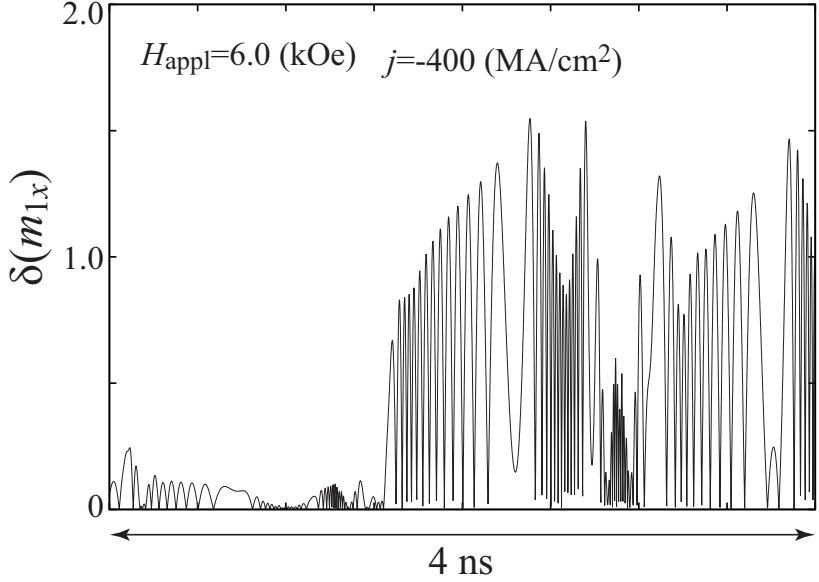}}
\caption{
        The magnitude of the difference between two lines in Fig. \ref{fig:fig9}(a). 
         \vspace{-3ex}}
\label{fig:fig10}
\end{figure}



A conventional approach to identify chaos in the field of nonlinear science is to evaluate Lyapunov exponent \cite{strogatz01}. 
The Lyapunov exponent $\lambda$ is defined as $|\delta(t)| \sim |\delta(0)| e^{\lambda t}$, 
where $\delta$ is the difference between two trajectories having slightly different initial conditions. 
We should note, however, that this definition of the Lyapunov exponent usually used cannot directly be applied to the present system. 
This is because the LLG equation conserves the norm of the magnetization. 
Therefore, for example, the range of the value of $m_{1x}$ in Fig. \ref{fig:fig9}(a) is restricted to $-1 \le m_{1x} \le +1$. 
This fact means that the difference between two lines in Fig. \ref{fig:fig9}(a) does not diverge monotonically; 
see Fig. \ref{fig:fig10}, where $\delta(m_{1x})$ is the magnitude of the difference between two trajectories shown in Fig. \ref{fig:fig9}(a). 
Such a non-monotonic motion cannot be described by a model in the form of $|\delta(t)| \sim |\delta(t)| e^{\lambda t}$. 
Thus, we consider that investigating the sensitive dependence on the initial condition is a reasonable approach to identify chaos in the present system. 



\section{Relaxation phenomenon}
\label{sec:AppendixC}

\begin{figure}
\centerline{\includegraphics[width=0.8\columnwidth]{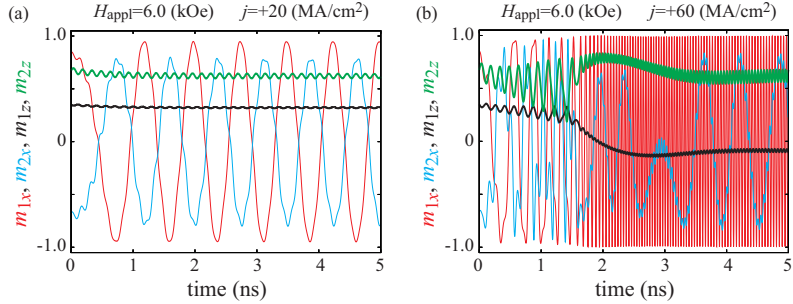}}
\caption{
         Time evolutions of $m_{1x}$ (red), $m_{2x}$ (blue), $m_{1z}$ (black), and $m_{2z}$ (green) near the initial states 
         for (a) $j=+20.0\times 10^{6}$ A/cm${}^{2}$ and (b) $j=+60.0 \times 10^{6}$ A/cm${}^{2}$. 
         The field magnitude is $H_{\rm appl}=6.0$ kOe. 
         \vspace{-3ex}}
\label{fig:fig11}
\end{figure}

Figures \ref{fig:fig11}(a) and \ref{fig:fig11}(b) show time evolutions of 
$m_{1x}$ (red), $m_{2x}$ (blue), $m_{1z}$ (black), and $m_{2z}$ (green) near the initial states 
for (a) $j=+20.0\times 10^{6}$ A/cm${}^{2}$ and (b) $j=+60.0 \times 10^{6}$ A/cm${}^{2}$. 
The applied field is $H_{\rm appl}=6.0$ kOe. 
We remind the readers that corresponding auto-oscillation states are shown in Figs. \ref{fig:fig3}(a), \ref{fig:fig3}(b), \ref{fig:fig4}(a), and \ref{fig:fig4}(b), 
i.e., \ref{fig:fig11}(a) shows the relaxation to a synchronization, whereas Fig. \ref{fig:fig11}(b) shows the relaxation to the oscillations with different frequencies. 
Figure \ref{fig:fig11} clarifies that the relaxation time to an auto-oscillation state is on the order of a few nanoseconds. 
Such a fast relaxation is suitable for applications such as MAMR.




It is preferable to clarify the relations among the material parameters, external forces (applied field and current), and relaxation time to auto-oscillation states 
from the perspectives of both fundamental and applied physics. 
However, it is not easy to answer this question due to the following reason. 
It is known that a small amplitude oscillation of the magnetization, i.e., ferromagnetic resonance (FMR), is well described by the linearized LLG equation \cite{vonsovskii66}. 
In fact, the FMR frequency is an eigenvalue of the linearized LLG equation. 
In this case, the magnetization oscillation and relaxation are described by an exponential function. 
Even for a nonlinear oscillation or synchronization phenomenon, described by Landau-Stuart and Adler equations respectively, 
the relaxation phenomena are described by complex forms of exponential functions \cite{pikovsky03}. 
On the other hand, it was shown that the relaxation phenomenon near the critical point is not described by an exponential function \cite{taniguchi17PRB}. 
We note that these cases are of special cases that can be solved in exact form. 
In general, in the presence of nonlinearity, as in the case of the LLG equation, the equation of motion can hardly be sovled exactly. 
Therefore, it is not clarified yet regarding what is the relation between material parameter and relaxation time nor what kind of function describes the relaxation phenomenon in STO.

Although past works on STO have focused on steady state properties such as emission power and linewidth \cite{kiselev03,rippard10,kubota13,suto12,hiramatsu16,slavin09}, 
the relaxation and/or transient phenomenon is becoming of great interest due to the following reasons. 
For example, in MAMR application, fast relaxation to auto-oscillation state is necessary to achieve fast recording. 
Another example is neuromorphic computing, where the computing is performed by using a sequence of input pulse data, 
where the history of input data is stored in the relaxation process of STO \cite{torrejon17}. 
Therefore, developing a comprehensive theory of relaxation phenomenon in STO will be highly required in future.








\end{document}